\documentclass[12pt, preprint]{aastex}

\begin{document}

\title{Flows and Non-thermal Velocities in Solar Active Regions
Observed with the Extreme-ultraviolet Imaging Spectrometer on
{\it Hinode}: A Tracer of Active Region Sources of Heliospheric
Magnetic Fields?}

\author{G. A. Doschek, H. P. Warren, J. T. Mariska}
\affil{Space Science Division, Naval Research Laboratory, Washington,
DC 20375, USA}
\email{george.doschek@nrl.navy.mil}

\author{K. Muglach}
\affil{ARTEP, Inc., Ellicott City, MD}

\author{J. L. Culhane}
\affil{Mullard Space Science Laboratory, University College London,
Holmbury St. Mary, Dorking, Surrey, RH5 6NT, UK}

\author{H. Hara, T. Watanabe} 
\affil{National Astronomical Observatory, Mitaka, Tokyo 181-8588}

\begin{abstract}

From Doppler velocity maps of active regions constructed from spectra
obtained by the Extreme-ultraviolet Imaging Spectrometer (EIS) on the
{\it Hinode} spacecraft we observe large areas of outflow (20-50 km
s$^{-1}$) that can persist for at least a day.  These outflows occur
in areas of active regions that are faint in coronal spectral lines
formed at typical quiet Sun and active region temperatures.  The
outflows are positively correlated with non-thermal velocities in
coronal plasmas.  The bulk mass motions and non-thermal velocities are
derived from spectral line centroids and line widths, mostly from a
strong line of \ion{Fe}{12} at 195.12 \AA.  The electron temperature
of the outflow regions estimated from an \ion{Fe}{13} to \ion{Fe}{12}
line intensity ratio is about $1.2-1.4\times10^6$ K.  The electron
density of the outflow regions derived from a density sensitive
intensity ratio of \ion{Fe}{12} lines is rather low for an active
region.  Most regions average around $7\times10^8$ cm$^{-3}$, but
there are variations on pixel spatial scales of about a factor of 4.
We discuss results in detail for two active regions observed by EIS.
Images of active regions in line intensity, line width, and line
centroid are obtained by rastering the regions.  We also discuss data
from the active regions obtained from other orbiting spacecraft that
support the conclusions obtained from analysis of the EIS spectra.
The locations of the flows in the active regions with respect to the
longitudinal photospheric magnetic fields suggest that these regions
might be tracers of long loops and/or open magnetic fields that extend
into the heliosphere, and thus the flows could possibly contribute
significantly to the solar wind.

\end{abstract}


\section{INTRODUCTION}

The launch of the Extreme-ultraviolet Imaging Spectrometer (EIS) on
the {\it Hinode} spacecraft on September 23, 2006 has given us the
opportunity to investigate the dynamics of coronal plasma, i.e, bulk
plasma flows and non-thermal motions, in more detail and with higher
spatial resolution than previously possible. In addition, the dynamics
can be associated with plasma diagnostic measurements of electron
temperature and density.  EIS can therefore help address fundamental
unsolved problems in active region physics, such as heating in active
region loops (e.g., Klimchuk 2006; Warren \& Winebarger 2007), and the
associations of active regions with the solar wind and heliospheric
magnetic flux (e.g., Liewer, Neugebauer, \& Zurbuchen 2004; Schrijver
\& DeRosa 2003).

Plasma motions might be particularly useful as tracers of either
active region heating or sources of solar wind and heliospheric flux.
For example, motions that are highly transient, spatially
well-localized and comprised of both upflows and downflows, might
signify active region loop heating and wave activity.  In contrast,
motions which are uni-directional (either upflows or downflows),
lasting over several days, and that cover an extensive spatial region
might be expected to be associated with primarily long loops or open
field regions, and are possible indicators of sufficiently open
magnetic flux to extend into the heliosphere and thus the flows could
contribute to the solar wind.  The association of flows with closed or
open field can be facilitated by the creation of flow maps of active
regions contructed by rastering the EIS spectrometer slit over an
active region, and then correlating the maps with the underlying
photospheric magnetic field.

Obtaining rasters of active regions and the creation of Doppler
velocity flow maps was an early science goal of the EIS data analysis
program.  Doschek et al. (2007) (and more recently Harra et al. 2008)
showed that in active regions there are extensive areas of low
intensity that are characterized by net bulk plasma outflows and
enhanced non-thermal motions.  The appearance of these outflow regions
in areas of very low intensity is striking when compared with maps of
spectral line intensity, also constructed from the EIS rasters.  In
this paper we expand the results of Doschek et al. (2007) and discuss
in detail EIS rasters of Doppler shifts and non-thermal motions for
two active regions.  We estimate electron temperatures and densities
of the outflowing regions, discuss data from other {\it Hinode}
instruments and other satellites, and speculate on the types of
magnetic structures that contain the flows and turbulent motions.  We
also estimate the mass flux of the flows into the corona.  We will
show that the locations of the flows in the active regions and the
underlying morphology of the photospheric longitudinal magnetic flux
suggest a link with heliospheric open fields that originate in active
regions.  The flows seem more associated with the solar wind and
heliospheric magnetic fields than with heating of active region loops.

Experimentally, flows are determined from measurements of a spectral
line centroid, coupled with information on the rest wavelength of the
line.  There are numerous observations of Doppler shifts in the
literature, the most recent being obtained from the Solar Ultraviolet
Measurements of Emitted Radiation (SUMER) spectrometer on the {\it
Solar and Heliospheric Observatory} (e.g., Teriaca et al. 1999a;
Warren, Mariska, \& Wilhelm 1997).  Mass motions can be related to
heating in coronal flux tubes and the acceleration of the solar wind
(e.g., Hassler et al. 1999; Tu et al. 2005; McIntosh et al. 2006).
The solar wind work has centered mostly on Doppler shift measurements
in coronal holes.  There are also numerous coronal flux tube dynamical
numerical simulation models in the literature (e.g., Hansteen 1993;
Warren \& Winebarger 2007; and Patsourakos \& Klimchuk 2006).  These
models attempt to explain observed density and temperature
measurements in coronal loops and relate the results to coronal
heating mechanisms.

The origin of excess optically thin spectral line widths beyond their
thermal Doppler widths is unclear.  Since a rocket flight by Boland et
al. (1975) it has been known that spectral lines from ions present in
the solar transition region and corona have wider profiles than
expected from pure thermal Doppler broadening alone, based on the
ionization equilibrium temperatures at which they are formed.  The
profiles of the majority of the lines are Gaussian or close to
Gaussian, and the excess broadening does not appear to be a strong
function of position on the solar disk.  The bulk of our current
knowledge of these motions in non-flaring plasma has been obtained
from the S082-B slit spectrograph on {\it Skylab} (e.g., Mariska
1992), the High Resolution Telescope Spectrograph (HRTS) rocket
stigmatic spectrograph (e.g., Bartoe 1982), and SUMER (e.g., Chae,
Sch\"uhle, \& Lemaire 1998a).

The excess spectral line broadening is usually expressed as a
non-thermal random mass motion component of unknown origin. It has
been cited as an expected signature of waves in the corona (e.g.,
Mariska, Feldman, \& Doschek 1978) and magnetic reconnection (e.g.,
Parker 1988).  It could also be due to different multiple plasma flow
speeds in structures smaller than the spatial resolution of the
spectrometer because the convolved Gaussian-like line profile would be
wider than the widths of the individual Gaussians composing the
overall profile.  Most recently, McIntosh et al. (2008) argue from
{\it Hinode} high spatial resolution Solar Optical Telescope (SOT)
observations that Alfv\,{e}nic waves observed in small-scale
spicular-type structures are responsible for the excess broadening.

Although there are numerous published observations of Doppler shifts
and non-thermal motions, there is little information, particularly for
the corona, on the detailed relationship of the motions to actual
structures, e.g., coronal loops.  There is far less information on the
relationship of these dynamical parameters to other physical
parameters such as electron temperature and density.  The apparent
close relationship between large line widths and bulk plasma flows
seen in the EIS spectra is one of the most interesting aspects of the
new observations.

In Section 2 we present pertinent details of the EIS on {\it Hinode}.
In Section 3 we discuss the observations and data reduction.  Results
are given in Section 4 and the implications of the results are
discussed in section 5.

\section{THE EIS ON {\it HINODE}}

The EIS is described in detail by Culhane et al. (2007) and Korendyke
et al. (2006).  The instrument consists of a multi-layer telescope and
spectrometer.  The telescope mirror and spectrometer grating are
divided into two halves, each of which is coated with different Mo/Si
multi-layers.  This results in two observed extreme-ultraviolet narrow
wavebands: 170-210 \AA\ and 250-290 \AA.  Light from the telescope is
focused onto the entrance aperture of the spectrometer.  The grating
then diffracts and focuses the light onto two CCD detectors.

The telescope mirror is articulated and different regions of the Sun
can be focused onto the spectrometer aperture by fine mirror motions.
The entrance aperture of the spectrometer can be one of four options:
a 1\arcsec\ slit, a 2\arcsec\ slit, a 40\arcsec\ slot, or a
266\arcsec\ slot.  The slits/slots are oriented in the solar
north/south direction.  Their heights can be variable, with a maximum
height for most observations of 512\arcsec.

There are several modes in which EIS can be operated.  Images of solar
regions can be constructed by rastering a slit or slot west to east
across a given solar area with a set exposure time at each step.  At
each raster position it is possible to read-out the entire CCD and
obtain a complete spectrum for each wavelength band.  It is also
possible to select a small set of lines, falling in narrower spectral
windows, the choices of which depend on the objectives of the
observation.

The spatial resolution of EIS along the slit is approximately
2\arcsec\ (1\arcsec\ per pixel) and the spectral dispersion is quite
high, 0.0223 \AA\ per pixel.  The instrumental full width at half
maximum measured in the laboratory prior to launch is 1.956 pixels.
However, we adopt an in-orbit instrumental width of 2.5 pixels, or
0.056 \AA.  This width was obtained by comparing the width of the EIS
\ion{Fe}{12} 193.51 \AA\ line observed above the limb with
\ion{Fe}{12} forbidden line observations made from the {\it Skylab}
S082-B spectrograph (Cheng, Doschek, \& Feldman 1979) and SUMER
(Doschek \& Feldman 2000).  The in-orbit instrument width is still
being investigated, and may undergo small revisions in the future.

\section{OBSERVATIONS AND DATA REDUCTION}

Data from two active regions were examined for this work.  These
regions are listed in Table 1, along with the start times of the EIS
observations, the solar locations, and the EIS exposure times.  The
locations are given in arcseconds relative to Sun center.  Positive
locations are west/north; negative ones are east/south.  The
observations consist of rasters using the 1\arcsec\ slit stepped from
west to east in 1\arcsec\ increments for a total of 255 pointings.
The slit height is 256 pixels.  Each spectrum at each location in the
raster contains 20 spectral lines, in 16 pixel wide spectral windows.
The locations in Table 1 refer to the centers of the rasters.  Data
from other active regions have been qualitatively examined and support
the general conclusions from the two regions discussed herein.

Most of the line width and position results discussed in this paper
were obtained from the 195.12 \AA\ \ion{Fe}{12} line.  This is the
most intense line in the EIS spectrum in part because its wavelength
is near the maximum of the multi-layer efficiency.  It is therefore
ideal for line profile measurements because good count rates are
available for statistically meaningful results.  In addition, the
ratio of \ion{Fe}{12} lines (186.89+186.85)/195.12) is used to obtain
electron densities, and the ratio of an \ion{Fe}{13} 202.04 \AA\ line
to the \ion{Fe}{12} 195.12 \AA\ line is used to detect changes in
electron temperature.  The atomic data were obtained from the CHIANTI
(version 5.2) atomic database (e.g., Landi et al. 2006).  All lines
discussed in this paper are listed in Table 2 with their temperatures
of peak formation in ionization equilibrium (Mazzotta et al. 1998).

Since EIS scans from west to east, the western edges of the active
regions were observed earlier than the eastern edges.  However, all
spectral quantities for a given spectral line (e.g., the parameters of
a Gaussian fit) compared at any single position in a raster were
recorded simultaneously on a single CCD.  Therefore, all spectral
quantities for a single line are perfectly co-aligned in position as
well as being recorded at the same time.

Observationally we have found that the width of the 195 \AA\ line is
slightly larger than the widths of the other \ion{Fe}{12} lines for
reasons not yet completely understood.  Part of the reason is a blend
with another relatively weak \ion{Fe}{12} line.  We have quantified
the difference by carrying out a detailed comparison between the
\ion{Fe}{12} 193.51 \AA\ and 195 \AA\ line widths using data from
another active region, from which a width correction factor for the
195 \AA\ line was determined.  We have linearly subtracted this
correction, 2.605 m\AA, from the 195 \AA\ width in order to derive
non-thermal velocities from the 195 \AA\ line.  Even without this
correction, we also found that we obtain the same qualitative
non-thermal line widths with the \ion{Fe}{13} line at 202 \AA\ as with
the 195 \AA\ line, as well as with the \ion{Fe}{12} line near 193 \AA.
We prefer the stronger 195 \AA\ line rather than the \ion{Fe}{12} 193
\AA\ line for line profile measurements because of its better counting
statistics.  Subtracting the 2.605 m\AA\ correction, compared with not
subtracting the correction, reduces the \ion{Fe}{12} non-thermal
velocity by about 5 km s$^{-1}$ at around 30 km s$^{-1}$ and by about
3 km s$^{-1}$ at about 70 km s$^{-1}$.

We have reduced the data using the standard EIS software data
reduction package.  The data were corrected for cosmic ray hits, hot
and warm pixels, detector bias, and dark current.  Data numbers were
converted to intensities in ergs cm$^{-2}$ s$^{-1}$ sr$^{-1}$
\AA$^{-1}$.  An additional effect that was corrected for is a
variation of line position over the {\it Hinode} orbit due to
temperature variations in the spectrometer caused mainly by a varying
exposure to infrared heating from Earth.  This variation produces an
approximately sinusoidal variation of spectral line centroids.  It is
removed by averaging centroid positions over a length of slit for
which the underlying solar region is mostly a quiet Sun region without
any obvious transient activity.  Another correction that was applied
is a correction for a slight tilt of the slit on the CCDs, which
affects wavelengths along a north/south direction.  These corrections
allow us to make Doppler maps of active regions in spectral lines for
comparisons with the non-thermal velocities, provided rest wavelengths
can be determined.

The parameters discussed in this paper are total spectral line
intensities, spectral line widths, and spectral line wavelengths.
These parameters are determined from the reduced data using a standard
Gaussian fitting procedure for each spectral line.

EIS does not have an absolute wavelength calibration source.  The
wavelength scale for EIS was calibrated as described in Brown et
al. (2007).  However, in order to make active region Doppler maps, a
choice must be made for each active region of a spectral line
wavelength that we decide represents zero Doppler velocity.  For each
active region discussed herein, we have determined this wavelength by
averaging low intensity regions over large areas that are outside of
the active region.  These defined rest wavelengths are within about
0.002 \AA\ or 3 km s$^{-1}$ of each other.

Previous work with SUMER (e.g., Brekke, Hassler, \& Wilhelm 1997;
Chae, Yun, \& Poland 1998b; Teriaca, Banerjee, \& Doyle 1999b)
indicates that the average quiet Sun Doppler velocity at the
temperature of \ion{Fe}{12} is close to zero, or at most a few
kilometers per second.  There is some indication that a small velocity
for \ion{Fe}{12} would be a blueshift (Teriaca et al. 1999b), which
would make the flows we observe slightly larger than we report here.
But this needs confirmation.

The non-thermal speed is obtained from the full width at half maximum
(FWHM, in m\AA) of the line profile from the expression,

\begin{equation}
FWHM = 1.665\times10^3~\frac{\lambda}{c}~\sqrt{\frac{2kT}{M} +
V^2}~~~,
\end{equation}
where $\lambda$ is the wavelength (in \AA\ in this paper), $c$ is the
speed of light, $k$ is the Boltzmann constant, $T$ is the electron
temperature, and $M$ is the ion mass.  Equation (1) assumes that the
instrumental width (56 m\AA) has been removed from the line profile
and that all broadening mechanisms are Gaussian.  Furthermore, if
ionization equilibrium is assumed, there is the tacit assumption that
the electron and ion temperatures are equal.  These assumptions are
highly likely valid in the low corona where the EIS observations are
made, because electron densities at these altitudes in the corona are
on the order of 10$^8$ to several times 10$^9$ cm$^{-3}$.  At these
densities equilibration times between electrons and ions are very
short, and ionization and recombination processes are very rapid for
ions such as \ion{Fe}{12}.

\section{RESULTS}

\subsection{The August 23, 2007 Active Region}

Figure 1 shows a summary of the results for the August 23, 2007 active
region.  The upper left panel is an intensity plot for the
\ion{Fe}{12} 195.12 \AA\ line.  The middle and lower left panels show
the line centroid shift (colorbar units are m\AA) and FWHM of the line
(colorbar units in km s$^{-1}$), respectively.  The top right panel
shows a co-aligned {\it Hinode} SOT magnetogram of the area within the
boxed region.The middle and bottom right hand panels show the centroid
and FWHM respectively within the boxed region shown in the left hand
panels.

The striking aspect of Figure 1 is that the Doppler shifts and
non-thermal line widths are not only well-correlated, but that the
largest values for these quantities occur in regions where the
\ion{Fe}{12} line intensity is very low.  The \ion{Fe}{10} (see Table
2) image does show activity in the area of the flows and non-thermal
motions, but it is not spatially correlated well with the \ion{Fe}{12}
features.  For the region in Figure 1, the largest Doppler shift
speeds (outflows) are on the order of 15-20 km s$^{-1}$ and the
non-thermal speeds become as large as 55-60 km s$^{-1}$.  The SOT
magnetogram in Figure 1 indicates that the outflow regions occur over
fields of one dominant magnetic polarity, which may indicate either
open field lines or long closed loops.

The temperature and density distributions in the flow regions are
shown in Figure 2.  The upper two panels show the electron temperature
and density as a function of the FWHM of the \ion{Fe}{12} line for the
region within the box in the upper right panel of Figure 1.  The
temperature was determined from the \ion{Fe}{13} to \ion{Fe}{12} line
ratio assuming ionization equilibrium and an isothermal plasma.  The
density was determined from the ratio of the \ion{Fe}{12} lines in
Table 2.  Figure 1 shows that the FWHM and Doppler shifts are
well-correlated, so the data in the upper panels of Figure 2 that
correspond to the largest FWHMs also correspond to the largest outflow
speeds.  There is a clump of data near a FWHM of 70 m\AA\ and a
scattering of points towards larger FWHM that does not either increase
or decrease with temperature.  These data arise from the outflow
regions.  The large outflow regions have a characteristic temperature
of about $1.2\times10^6$ K and a density of about $7\times10^8$
cm$^{-3}$, just above quiet Sun values.  Thus the large outflow
regions are more representative of quiet Sun regions than active
regions.

The bottom two panels of Figure 2 show histograms of the Doppler shift
and the FWHM.  Both exhibit tails that show the correlation of FWHM
with Doppler speed.  The data in Figure 2 that clump near 70 m\AA\ and
near zero Doppler speed are from the pixels surrounding the outflow
regions.  The clear correlation of Doppler speed and non-thermal
velocity is shown in Figure 3.  These results were obtained by
removing the instrumental width from the line profiles and assuming an
electron temperature as defined for Figure 2.

The August 23 active region was not bright enough to compare with good
statistics the \ion{Fe}{12} outflow speeds and FWHMs with other lines.
The strength of the activity was weak enough such that there is no
visible sunspot at the location of the active region.  However, there
is a general positive correlation with similar data from the
\ion{Fe}{13} line in Table 2, and some, but reduced, correlation with
the \ion{Fe}{10} line.  {\it Transition Region and Coronal Explorer}
({\it TRACE}) images show considerable activity in the 171 \AA\ band
near the EIS \ion{Fe}{12} outflows, but the spatial correlation of
{\it TRACE} and EIS features is not particularly good.

The spectral line profiles in the wide-line outflow regions look
fairly symmetric, with no obvious blueshifted separate component to an
otherwise stationary component, as seen in X-ray line profiles at the
onset of many flares. This result is similar to that found by Doschek
et al. (2007) for an active region observed in December, 2006.

We note that there was considerable activity in the bright part of the
August 23 region throughout the time of the EIS raster.  Observations
with {\it TRACE} and the Extreme UltraViolet Imagers (EUVI) on the two
{\it Solar Terrestrial Relations Observatory} ({\it STEREO})
spacecraft show that the obvious large loop system in the shape of a
$u$ in the \ion{Fe}{12} intensity image near the center of the image
did not exist at the start time of the EIS raster.  It is the result
of a filament eruption that occurred with the ends of the filament
apparently anchored east and west of the outflow regions.  The outflow
regions do not appear to be involved in any of the activity to the
east in the bright part of the active region.

The August 23 region was also observed on August 22, twice during
August 23, and once on August 24.  The August 22/23 observations cover
a 24 hour period and the August 24 observation extends the coverage by
about 12 more hours.  Outflows in the same general area are seen as in
the observation under discussion.  Thus the outflows appear to be
persistent with lifetimes on the order of at least 1.5 days.  This
result was also found by Del Zanna (2008), who in addition confirmed
the conclusions in Doschek et al. (2007).

\subsection{The December 11, 2007 Active Region}

Figure 4 shows results for the December 11 active region presented in
the same format as in Figure 1.  Again, there is a clear separation of
the largest outflow regions from the most intense regions in the
\ion{Fe}{12} line, and there is a clear correlation between outflows
and spectral line widths.  And again, the large outflow regions are
located in an area of primarily a single polarity.  However, the
December 11 active region is stronger than the August 23 region and
contains a prominent sunspot group.

The SOT G band, Ca H, and magnetogram images of the active region are
shown in Figure 5, co-aligned with the EIS FWHM image.  It is clear
that the large sunspot in the G band image lies south of the large
FWHM region.  The circular region surrounded by large FWHM regions in
the upper left Figure 5 panel is not delineating the sunspot, but
rather is close to a region of weak magnetic field, as seen by
comparing the magnetogram with the FWHM image.  This circular region
is also a region of weak Ca H emission.  The alignment of the region
in the SOT data with the EIS data is not precise, but is within about
2\arcsec.

Similar results as shown for the August 23 region in Figure 2 are
shown for the December 11 region in Figure 6.  These results
correspond to the data within the boxed area shown in the upper right
panel of Figure 4.  As found for the August 23 region, pixels for
which large FWHM values are derived are characterized by quiet Sun
temperatures and relatively low densities.  In contrast, the regions
immediately surrounding the large width areas have densities in places
that are on the order of $10^{10}$ cm$^{-3}$ and somewhat higher
electron temperatures.

The lower two panels of Figure 6 show histograms of the Doppler shift
and FWHM for the \ion{Fe}{12} line.  In this case there is clearly a
strong outflow component in the Doppler shift histogram, much stronger
than for the August 23 region.  Although the color scheme in Figure 4
shows where the strongest outflows are coming from, this can be seen
more clearly from Figure 7.  The upper left panel of Figure 7 shows
the Doppler shift, and the black and white images show where Doppler
shifts of different magnitudes come from.  The white pixels in each
image represent those pixels for which the Doppler shift shown in
Figure 6 is equal to or less than the value of $Ds$ m\AA\ specified
for each image.  Thus the data in the upper right panel of Figure 7
illustrate that all the outflows pixels with Doppler shifts less than
-8 m\AA\ occupy the white area shown.  The position of this value of
$Ds$ can be seen in the Doppler shift histogram of Figure 6 and it
represents essentially all of the outflow regions.  The part of the
Doppler shift histogram near zero shift in Figure 6 arises from all
the regions surrounding the dark region in the upper left panel of
Figure 7.  As $Ds$ is progressively decreased, the white area is
reduced and reveals clearly the regions of strongest Doppler shift.
The shift $Ds$ can be converted to a speed $V$ in km s$^{-1}$ by, $V$
= $3\times10^{5}$$\times$$Ds$/195.120.

The correlation between non-thermal speed and Doppler shift is given
in Figure 8.  The non-thermal speeds reach values of about 90 km
s$^{-1}$ and the maximum Doppler outflow speeds are on the order of 45
km s$^{-1}$.  Note that Doppler shifts as large as 140 km s$^{-1}$, as
reported by Sakao et al. (2007) for a different active region, are not
observed (However, see Harra et al. 2008).

The December 11 active region is bright enough to examine the outflow
images in a number of different spectral lines (see Table 2), formed
at different electron temepratures.  Results for the December 11
outflow region are shown in Figure 9.  In Figure 9 there is a general
resemblance of all the outflow images, implying a correlation.
However, a close inspection shows that the easternmost (leftmost)
outflowing region has approximately the same location in all the
images, but the westernmost outflowing region shifts gradually more
westward as the temperature of line formation increases.  This is
clearly seen by comparing the rightmost upper and lower panels in the
figure, showing the \ion{Fe}{12} and \ion{Fe}{15} images,
respectively.  Note also that the fine structure varies among all the
images, implying small-scale multi-temperature outflowing regions.

\ion{Fe}{12} line profiles in the outflow regions are symmetric in
some areas, particularly the westernmost regions, but in the
easternmost regions an asymmetric blue wing appears frequently on
profiles.  This further supports the concept of multiple flow sites
with flows at different Doppler speeds.

Inspection of EIS images for spectral lines other than the
\ion{Fe}{12} line formed at lower and higher temperatures, as well as
{\it STEREO}/EUVI movies, shows many long and not too bright loops in
the vicinity of the outflows.  These loops generally have a westerly
component; some are even mostly parallel to the east/west direction.
Other loops have substantial southern and northern components.

\section{DISCUSSION}

From the above analysis of active region dynamics with EIS we may
conclude:

1. Extensive areas of active regions where the \ion{Fe}{12} emission
line intensity is weak exhibit higher Doppler shift outflows and
higher non-thermal velocities than found in other areas of the active
regions where the line intensity is much stronger.  This does not rule
out large flows and line widths in high intensity areas of active
regions; we are simply implying that there will be regions of low
intensity in active regions that have significant Doppler shift and
line width signals.  The outflows range from a few km s$^{-1}$ to as
much as 50 km s$^{-1}$.  The non-thermal motions range from about 20
to 90 km s$^{-1}$.

2. The temperatures of the outflow regions obtained from the intensity
ratio of an \ion{Fe}{13} line to an \ion{Fe}{12} line are about
$1.4\times10^6$ K, and electron densities vary from about
$5\times10^8$ to about $10^{10}$ cm$^{-3}$, depending on the
particular region.

3. The outflow regions are concentrated primarily over or near
magnetic regions of a single polarity, and can last for at least
periods of a day and a half although there are variations within the
flow region.

4. There is some evidence for variations of temperature within flow
structures and multiplicity of flows, i.e., the excess widths of the
lines need not be due to turbulence even in cases where the line
profiles appear well-fit by single Gaussian profiles.  It is possible
that a wide profile is composed of the line profiles of many
outflowing regions, each slightly Doppler shifted with respect to each
other due to slightly different Doppler shifts.  This scenario might
produce a summed wide profile that could be fit with a single
Gaussian.  A multi-component scenario definitely holds for the 11
December event.  Some of the outflow profiles show an obvious
secondary outflowing component such that the overall profile departs
significantly from a Gaussian.

5. The \ion{Fe}{12} intensity in the outflow regions is much less than
in the bright active region loops.  However, it is brighter than in a
disk coronal hole that was observed by EIS on September 28, 2007 near
01:45 UT.  In this coronal hole the \ion{Fe}{12} line is significantly
fainter than in the active region outflow regions.

The faintness of the structures could be due to intermittency in the
outflows.  That is, if the outflows are events with time scales
significantly less that the exposure times, they will appear faint
simply because they are bright for only part of the exposure time. For
example, two hypothetical regions, equally bright per second, would
appear different by a factor of, say, 5 in brightness if one region
only lasted 5 seconds while the other lasted for a longer exposure
time of 25 seconds.  Intermittent events of this nature would have to
persist over time intervals of at least a day to explain the
persistence of the outflows.

From inspection of context data from a number of spacecraft it appears
that the outflow regions are associated with long magnetic flux tubes,
and/or open field lines that extend into the heliosphere.  It is
interesting to calculate the mass flux in the outflowing regions. For
the December 11 event the outflow mass flux for all outflows greater
than about 12 km s$^{-1}$ in the boxed region of Figure 4 is about
$1\times10^{-5}$ gm cm$^{-2}$ s$^{-1}$ and is about $1.6\times10^{-6}$
g cm$^{-2}$ s$^{-1}$ for speeds exceeding 31 km s$^{-1}$.  For the
August 23 event the mass outflow flux in the boxed region of Figure 1
is about $5.7\times10^{-7}$ g cm$^{-2}$ s$^{-1}$ for speeds exceeding
12 km s$^{-1}$.  These quantities are obtained by multiplying the flow
speed by the mass density for each pixel in the boxed regions with a
outflow speed greater than the specified speed, and then summing the
result over all pixels.

More investigation is needed to determine whether or not this mass
flux, or a fraction of it, contributes to the solar wind.
Alternatively, some of the mass flux might be confined within long
closed flux tubes.  Liewer et al. (2004), Schrijver \& DeRosa (2003),
Luhmann et al. (2002), Neugebauer et al. (2002), and Wang \& Sheeley
(2002) have found from data and modeling that active regions can
contribute substantially to the heliospheric magnetic field and solar
wind.  Liewer et al. (2004) find that the open field lines are from
the edges of active regions (see Figures 11 and 12 in their paper), in
locations similar to where we find the flows discussed in this paper.
Schrijver \& DeRosa (2003) assert that connections with the
heliospheric field to active regions are common, long-lived, and that
during solar maximum the contributions of active regions to the
interplanetary magnetic field can be as much as 50\%.  As seen in
Liewer et al. (2004), Figure 14 in the Schrijver \& DeRosa paper shows
open field lines connected to the edge of a plage region.  Also, a
comparison of their simulation results with the {\it TRACE} image in
their figure shows that the open field regions emanate from a region
that is largely dark in the {\it TRACE} image, qualitatively similar
to the locations of the EIS flows in the active regions we have
discussed. It is therefore quite tempting to associate at least some
of the EIS flows as sources of the solar wind confined to open field
lines that extend into the heliosphere.

Some of the Doppler results of this paper are similar to those found
by Harra et al. (2008), who also analyzed flows in an active region
discussed previously by Sakao et al. (2007), and these authors suggest
that the flows might contribute to the solar wind.  These authors
found the same types of flows as reported earlier by Doschek et al
(2007), but Doschek et al. (2007) did not make a connection with the
solar wind.

In addition to the two active regions analyzed in detail in this
paper, we have qualitatively examined five other active regions that
exhibit similar behavior to the two regions we discussed in detail.
However, more studies with EIS will be needed along with simulations
of the solar magnetic field in order to test quantitatively the solar
wind connection.  We are continuing to obtain high resolution rasters
of active regions and are refining our knowledge of the zero velocity
centroid wavelengths of lines in the EIS spectrum.  In particular,
deeper exposures that reveal the flows in spectral lines formed over a
broad temperature range should be useful in connecting the flows back
to their loop footpoint origins.  More work also needs to be done on
the line profiles in some of the flow regions, which exhibit obvious
indications of multiple loop flows.

{\it Hinode} is a Japanese mission developed and launched by
ISAS/JAXA, collaborating with NAOJ as domestic partner, and NASA (USA)
and STFC (UK) as international partners.  Scientific operation of the
{\it Hinode} mission is conducted by the {\it Hinode} science team
organized at ISAS/JAXA.  This team mainly consists of scientists from
institutes in the partner countries.  Support for the post-launch
operation is provided by JAXA and NAOJ, STFC, NASA, ESA (European
Space Agency), and NSC (Norway).  We are grateful to the {\it Hinode}
team for all their efforts in the design, build, and operation of the
mission.

The authors acknowledge support from the NASA {\it Hinode} program and
from ONR/NRL 6.1 basic research funds.

\clearpage

\begin{table}\label{table1}
\caption{Observed Active Regions in 2007}
\begin{center}
\begin{tabular}{llll}
\hline
\hline
Date & Time (UT) & Location & EIS Exposure Time (s)\\
\hline
23~August & $01:55:43$ & $-518\arcsec,-211\arcsec$ & $60$\\
11~December & $00:24:16$ & $-178\arcsec,-144\arcsec$ & $60$\\ 
\hline
\end{tabular}
\end{center}
\end{table}

\begin{table}\label{table1}
\caption{Spectral Lines and Temperatures of Formation}
\begin{center}
\begin{tabular}{lll}
\hline
\hline
Ion & Wavelength (\AA) & Temperature (K)\\
\hline
\ion{Fe}{8} & $185.21$ & $6.3\times10^5$\\
\ion{Fe}{10} & $184.54$ & $1.0\times10^6$\\
\ion{Fe}{12} & $186.89+186.85$ & $1.6\times10^6$\\
\ion{Fe}{12} & $195.12$ & $1.6\times10^6$\\
\ion{Fe}{13} & $202.04$ & $1.6\times10^6$\\
\ion{Fe}{14} & $274.20$ & $1.8\times10^6$\\
\ion{Fe}{15} & $284.16$ & $2.0\times10^6$\\
\hline
\end{tabular}
\end{center}
\end{table}

\clearpage

\begin{figure*}[t!]
\centerline{
\includegraphics[clip,scale=0.7]{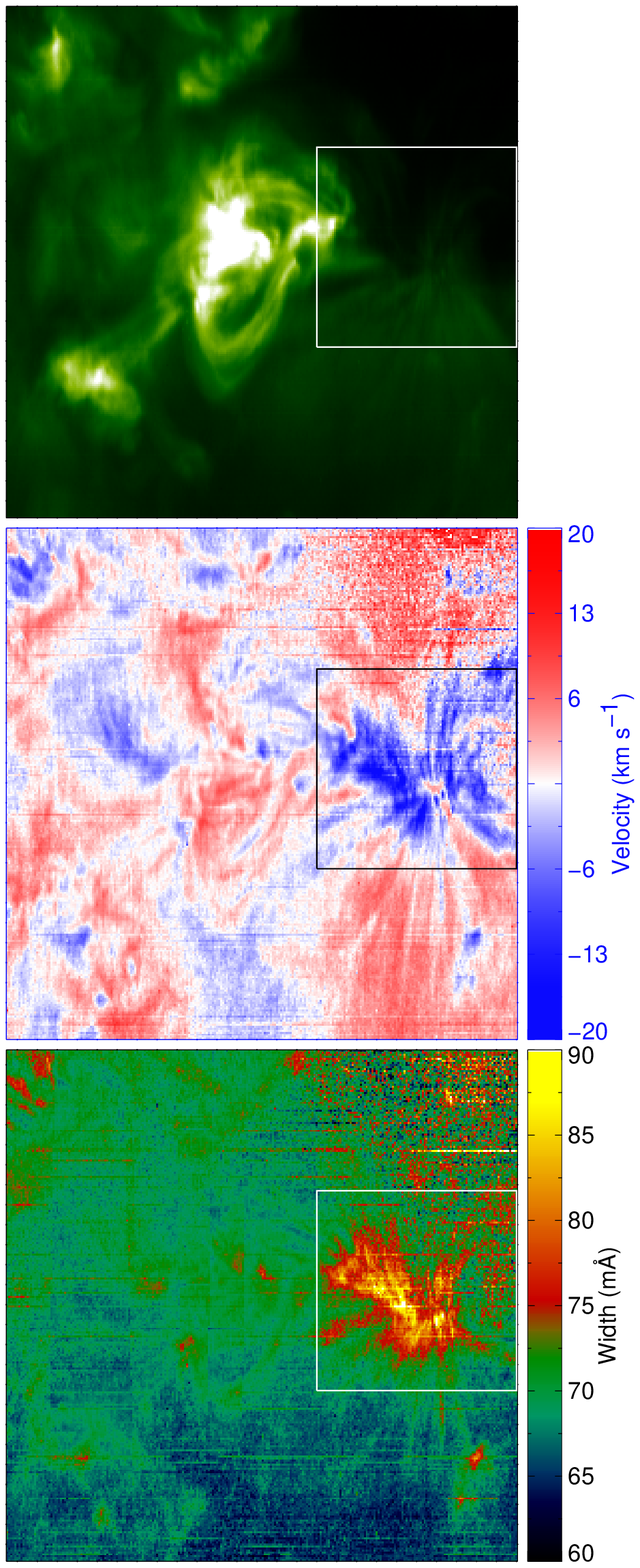}
\includegraphics[clip,scale=0.7]{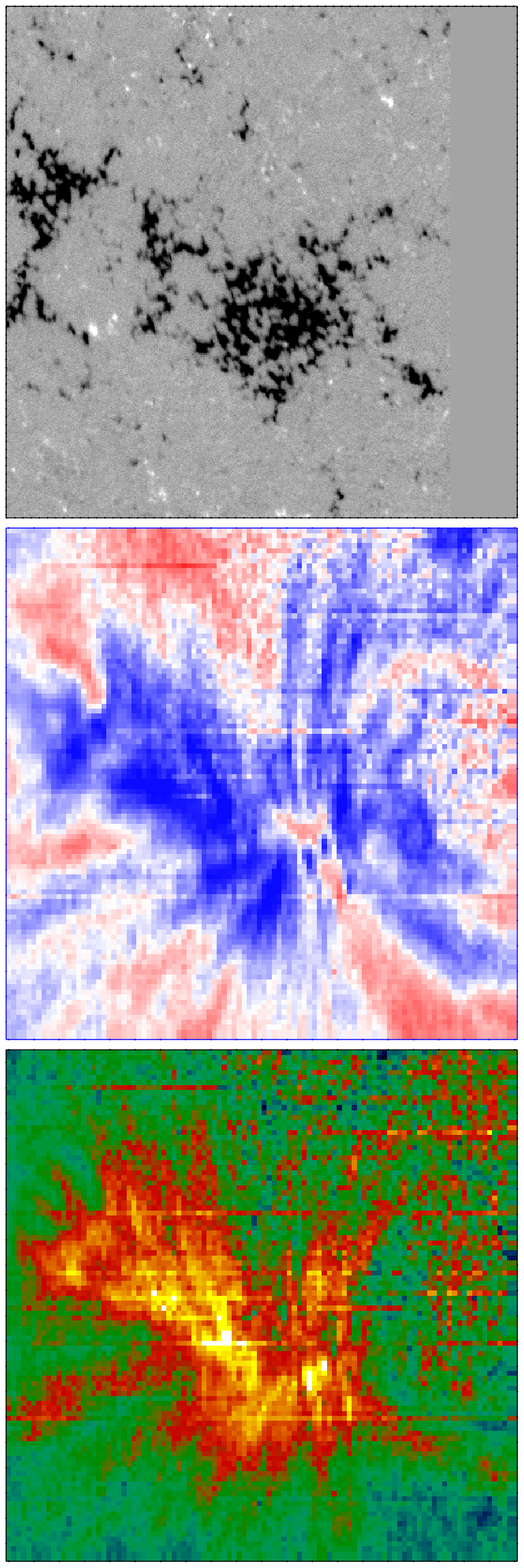}
}
\caption{Left panels: Top - images of \ion{Fe}{12} 195.12 \AA\
intensity for the August 23 active region; middle - centroid shift
(blue is towards the observer); bottom - line width; right panels: top
- SOT magnetogram within the boxed region; middle - centroid shift of
region in the boxed area in the left panels; bottom - line width
within the boxed region.  The ordinates are in the north-south
direction; the abscissae are in the east-west direction.}
\end{figure*}

\clearpage

\begin{figure}[t!]
\plotone{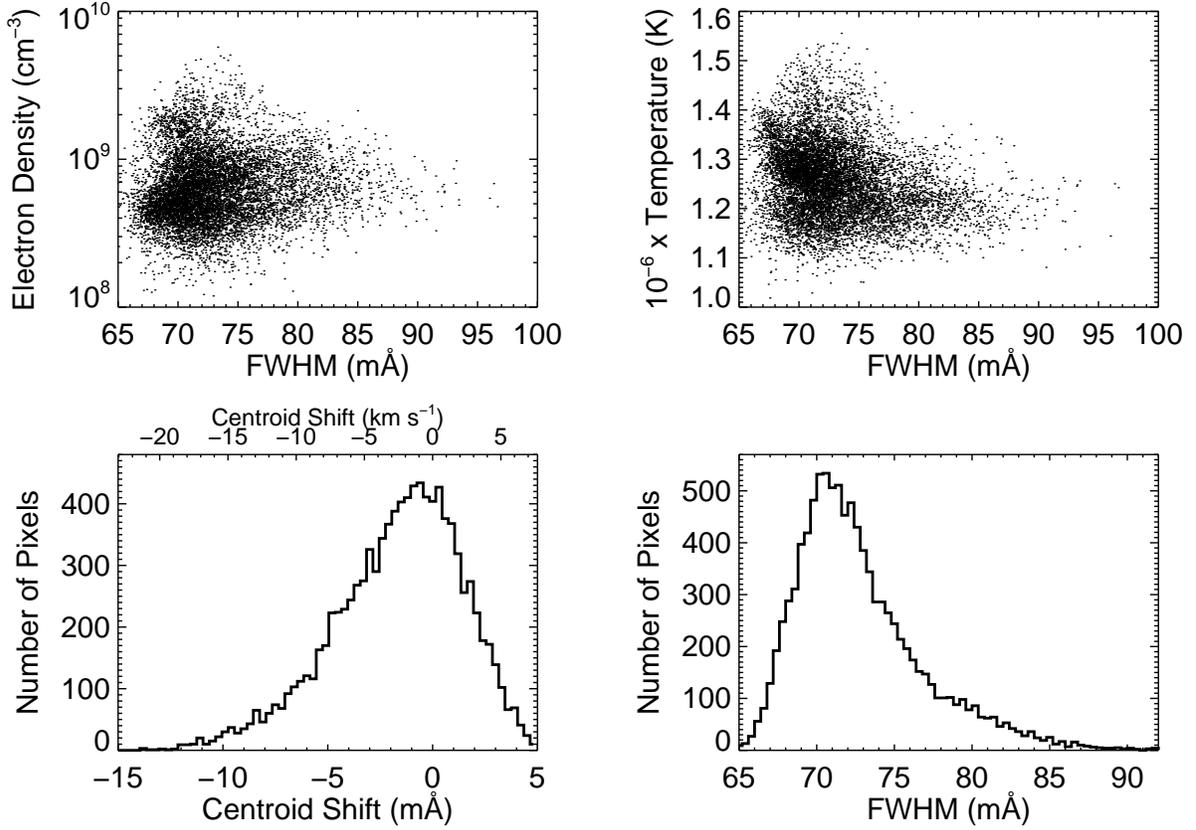}
\caption{Top panels: Electron density and temperature distributions
within the boxed area of Figure 1 as a function of FWHM of the Fe XII
195 \AA\ line.  The temperature is derived from the
\ion{Fe}{13}/\ion{Fe}{12} ratio.  Bottom panels: histograms of
centroid shift (left) and FWHM (right) within the boxed area of Figure
1.  The centroid shift in m\AA\ is converted to Doppler speed (upper
axis).  In the bottom right panel, 65 m\AA\ FWHM corresponds to 16 km
s$^{-1}$ non-thermal velocity and 92 m\AA\ corresponds to 61 km
s$^{-1}$.}
\end{figure}

\clearpage

\begin{figure}[t!]
\plotone{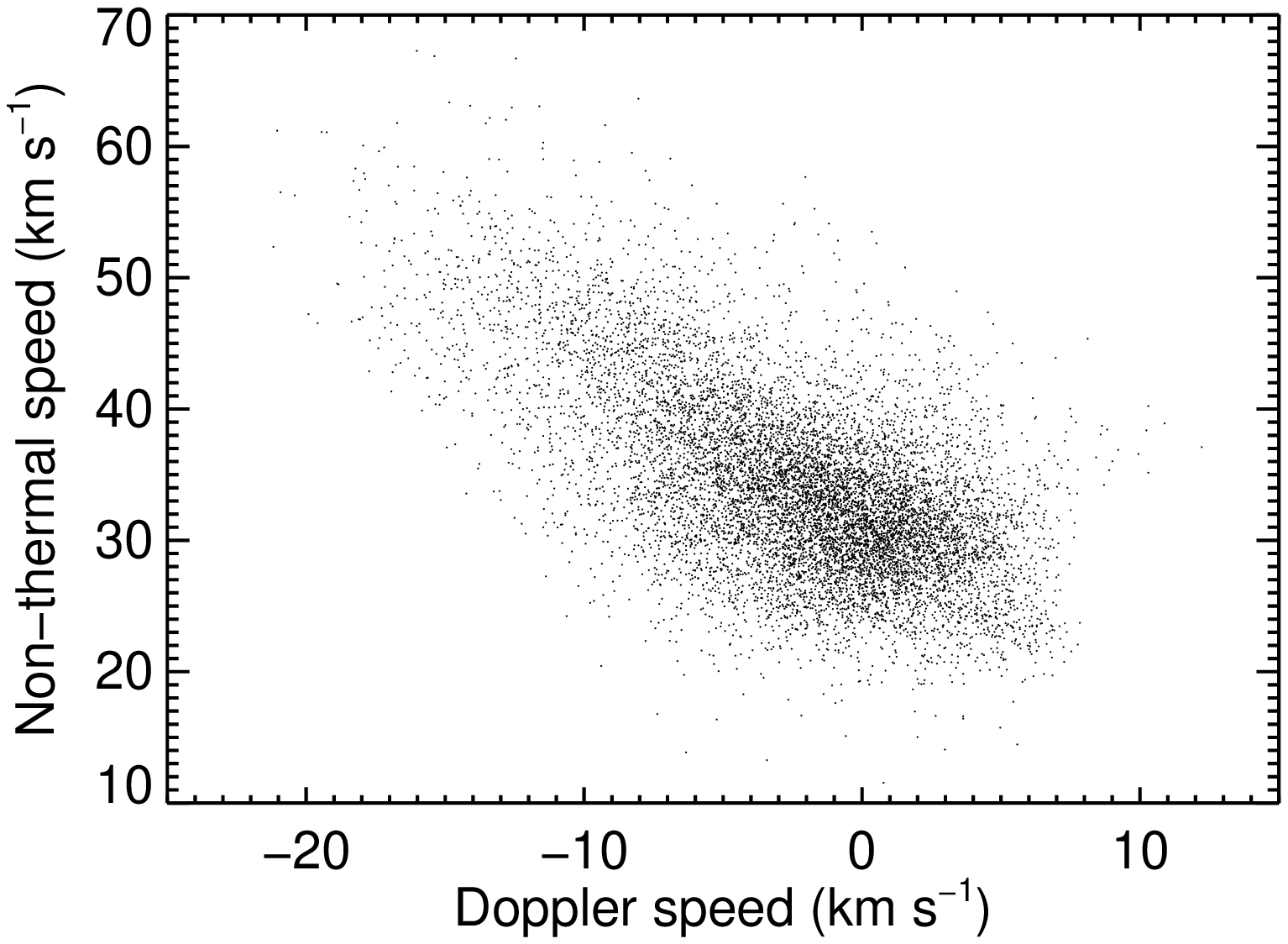}
\caption{Correlation of Doppler speed with non-thermal speed for the
data within the boxed area of Figure 1.}
\end{figure}

\clearpage

\begin{figure*}[t!]
\centerline{
\includegraphics[clip,scale=0.7]{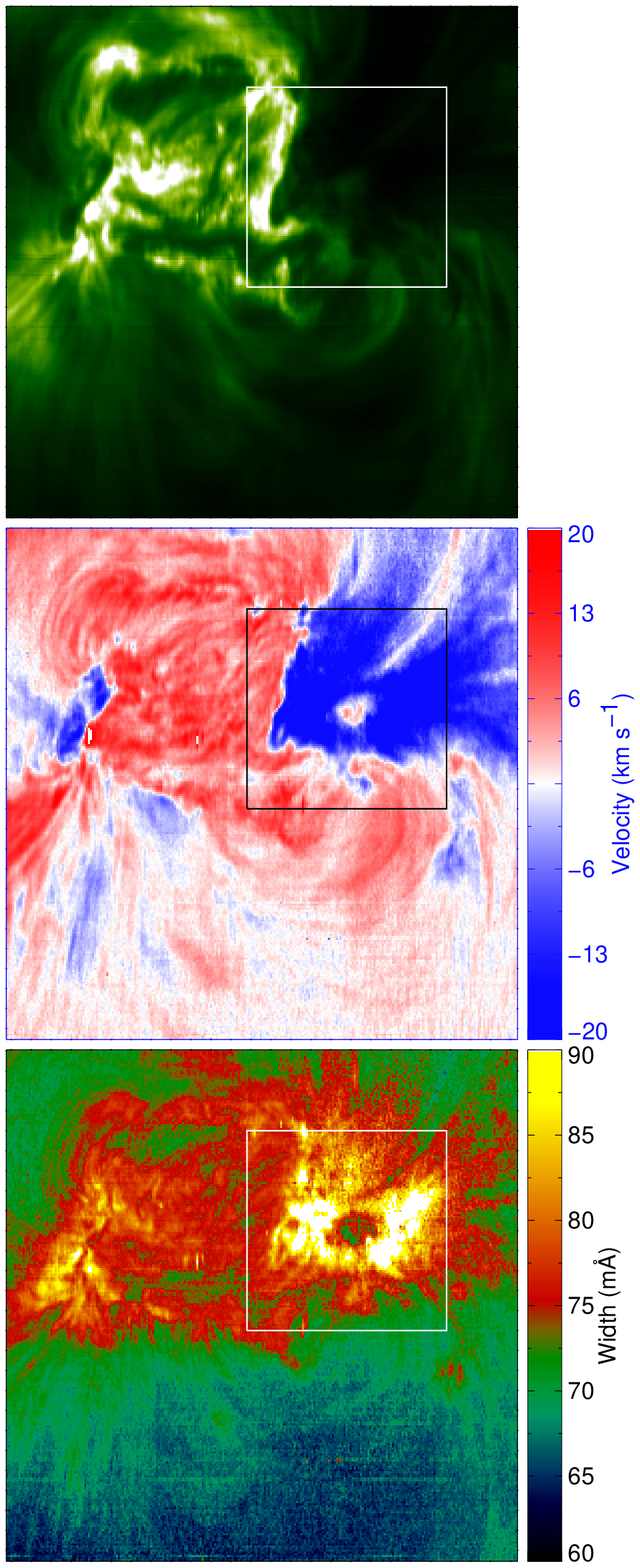}
\includegraphics[clip,scale=0.7]{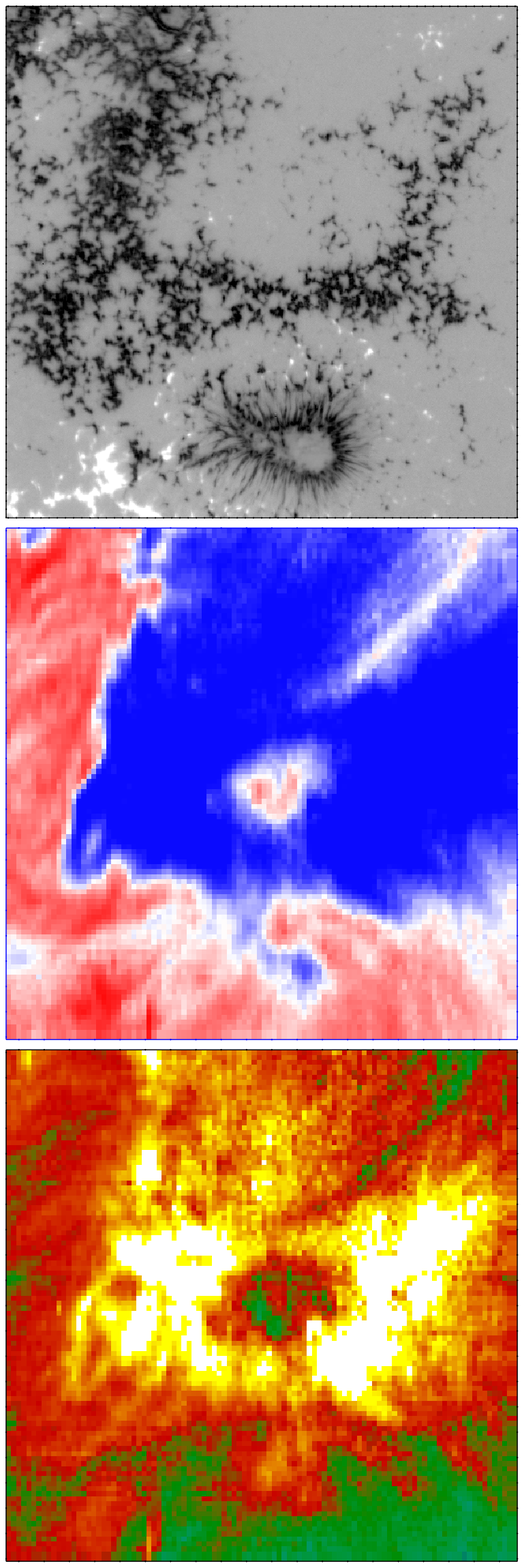}
}
\caption{Left panels: Top - images of \ion{Fe}{12} 195.12 \AA\
intensity for the December 11 active region; middle - centroid shift
(blue is towards the observer); bottom - line width; right panels: top
- SOT magnetogram within the boxed region; middle - centroid shift of
region in the boxed area in the left panels; bottom - line width
within the boxed region.  The ordinates are in the north-south
direction; the abscissae are in the east-west direction.}
\end{figure*}

\clearpage

\begin{figure}[t!]
\plotone{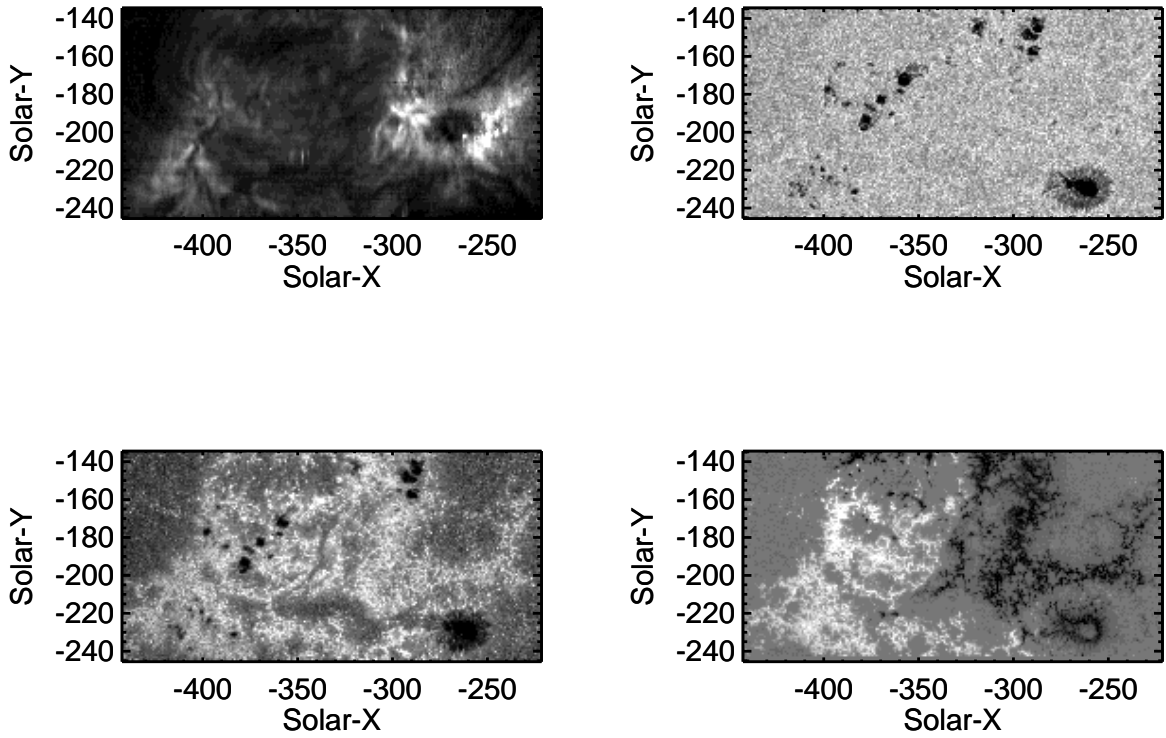}
\caption{Upper left and lower right panels: the \ion{Fe}{12} FWHM and
SOT magnetogram shown in Figure 4, respectively. Upper right panel:
the SOT G-band image; lower left panel: the SOT Ca H image.}
\end{figure}

\clearpage

\begin{figure}[t!]
\plotone{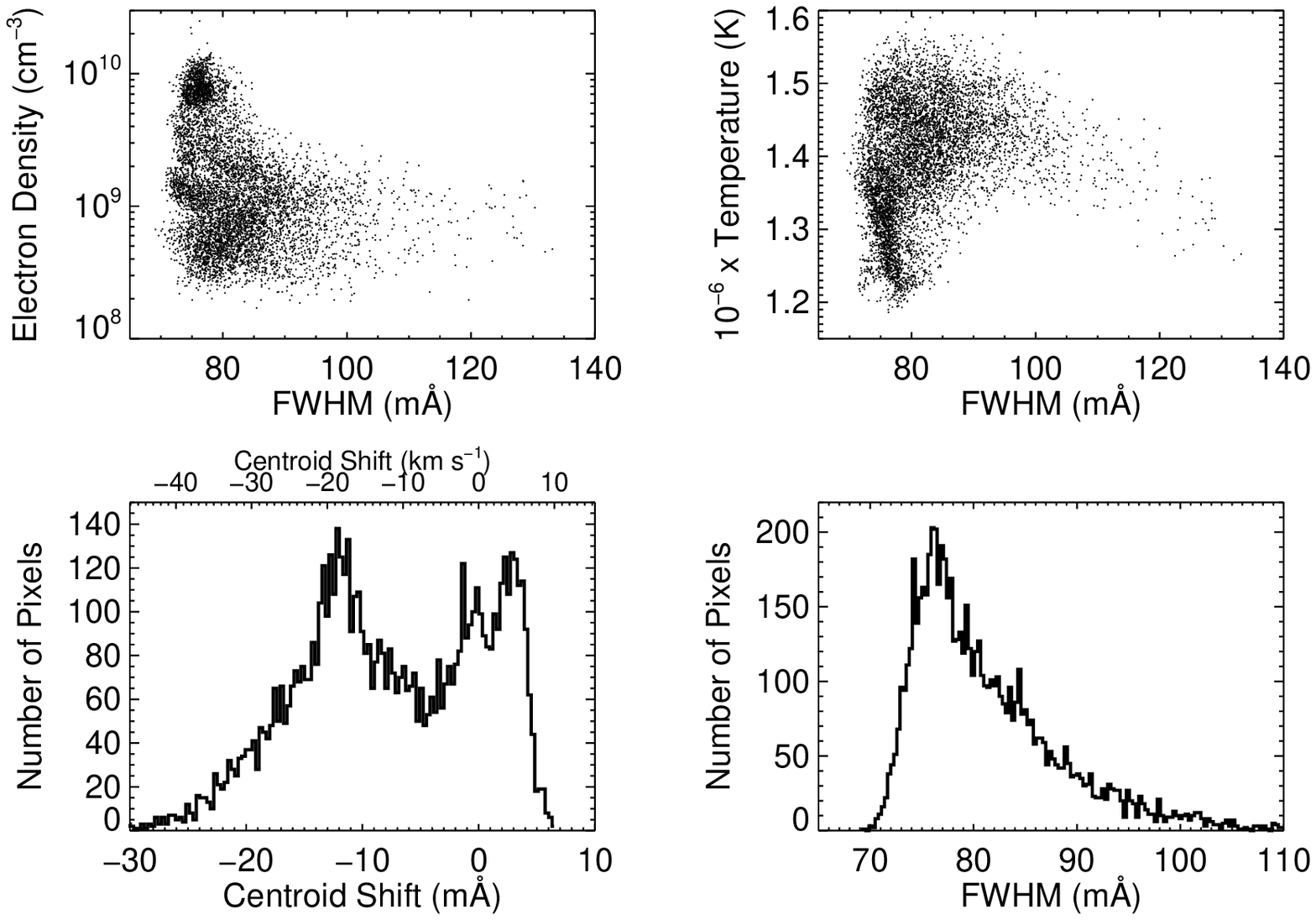}
\caption{Top panels: Electron density and temperature distributions
within the boxed area of Figure 4 as a function of FWHM of the Fe XII
195 \AA\ line.  The temperature is derived from the
\ion{Fe}{13}/\ion{Fe}{12} ratio.  Bottom panels: histograms of
centroid shift (left) and FWHM (right) within the boxed area of Figure
4.  The centroid shift in m\AA\ is converted to Doppler speed (upper
axis).  In the bottom right panel, 65 m\AA\ FWHM corresponds to 16 km
s$^{-1}$ non-thermal velocity and 110 m\AA\ corresponds to 82 km
s$^{-1}$.}
\end{figure}

\clearpage

\begin{figure}[t!]
\plotone{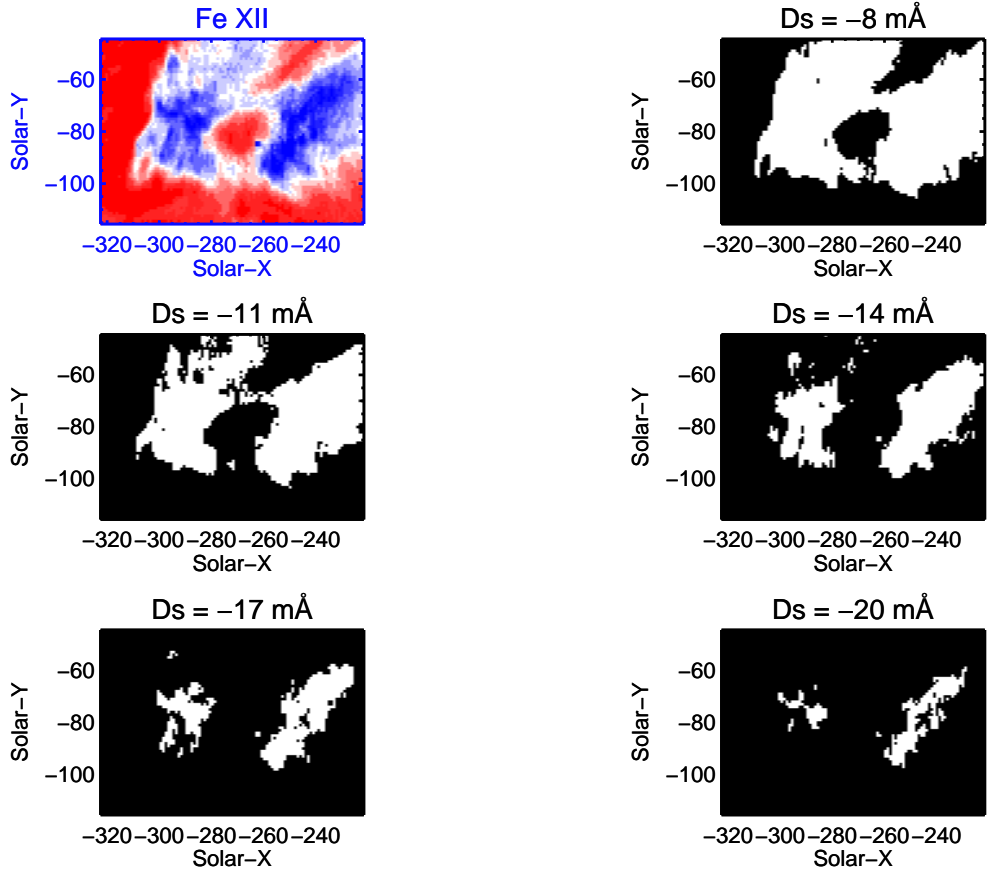}
\caption{Locations of outflows for the December 11 event (see Figure
4) according to magnitude of outflow.  The white areas in the black
and white figures represent pixels where the Doppler shift is equal
to or less than $Ds$ from zero Doppler shift in Angstrom units.
Negative $Ds$ is defined as an outflow.}
\end{figure}

\clearpage

\begin{figure}[t!]
\plotone{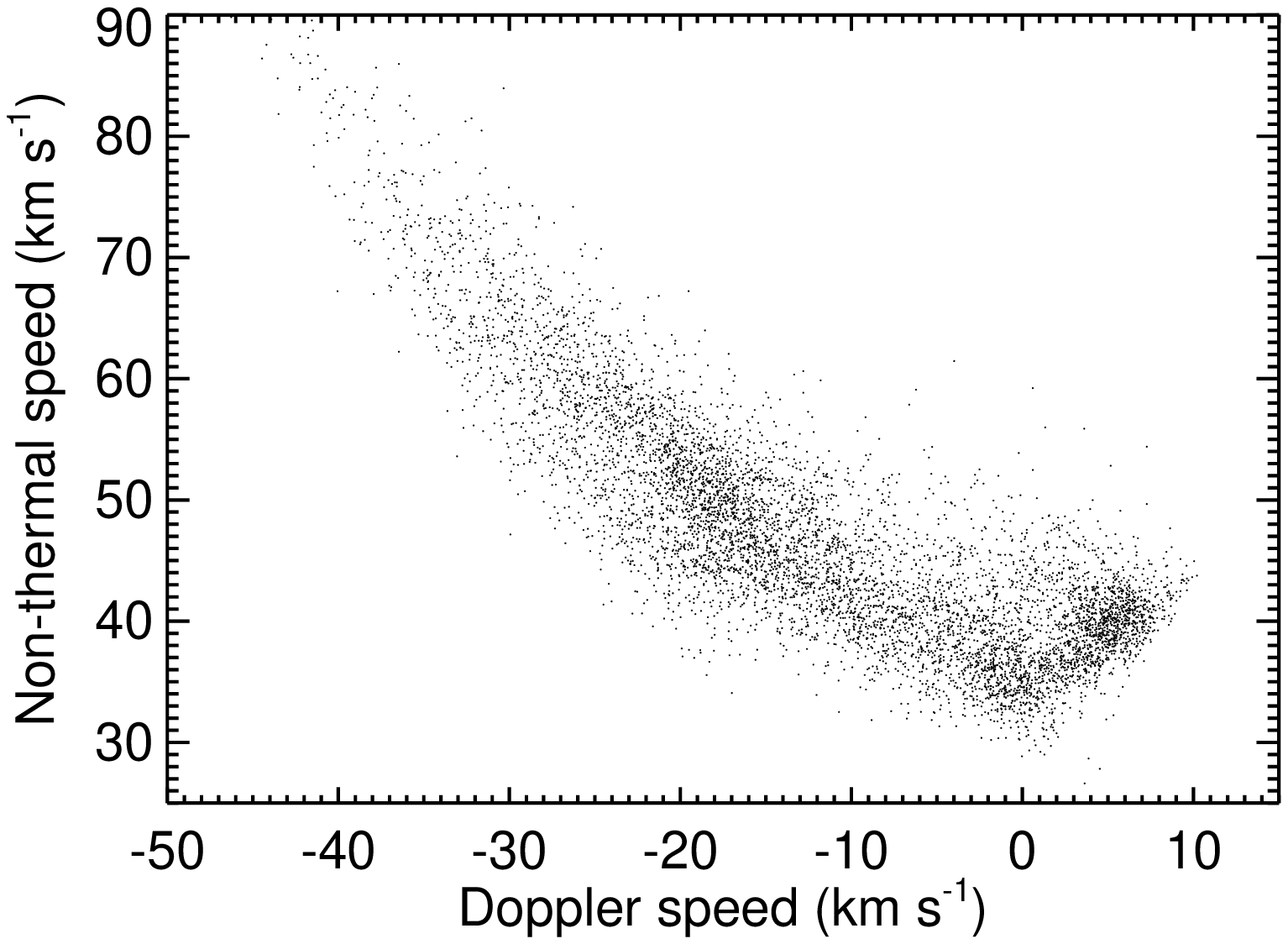}
\caption{Correlation of Doppler speed with non-thermal speed for the
data within the boxed area of Figure 4}
\end{figure}

\clearpage

\begin{figure}[t!]
\plotone{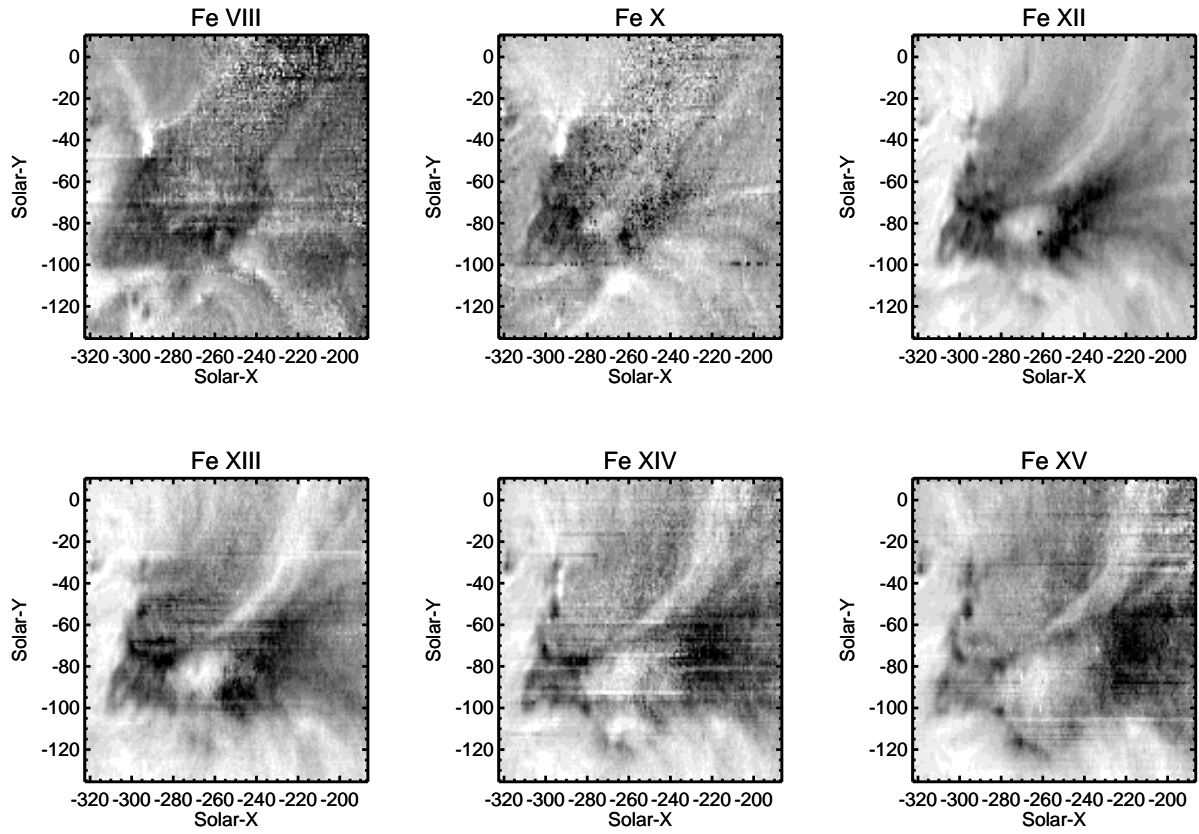}
\caption{Comparison of Doppler shifts interpreted as outflows for the
December 11 event (see Figure 4 boxed area) in different spectral
lines (see Table 2) formed over a range of temperatures.}
\end{figure}

\clearpage

\end{document}